\documentclass[twocolumn]{aastex62}

\usepackage{amsmath,amssymb,amsfonts}
\usepackage{subfigure}
\usepackage{natbib}
\usepackage{times}
\usepackage{booktabs}

\newcommand{\rv}[1]{{#1}}
\newcommand{\rvv}[1]{\textbf{#1}}
\begin{document}
\title{Simulation of Alfv\'en wave propagation in magnetic chromosphere with radiative loss: effects of non-linear mode coupling on chromospheric heating}

\author{Yikang Wang}
\affil{Department of Earth and Planetary Science, The University of Tokyo, Tokyo 113-0033, Japan}
\author{Takaaki Yokoyama}
\affil{Department of Earth and Planetary Science, The University of Tokyo, Tokyo 113-0033, Japan}

\email{wangyk@eps.s.u-tokyo.ac.jp}

\begin{abstract}
 We perform magnetohydrodynamic (MHD) simulations to investigate the propagation of Alfv\'en wave in magnetic chromosphere. We use the 1.5-dimensional expanding flux tube geometry setting and transverse perturbation at the bottom to generate the Alfv\'en wave. Compared with previous studies, our expansion is that we include the radiative loss term introduced by \cite{2012A&A...539A..39C}. We find that when an observation based transverse wave generator is applied, the spatial distribution of the time-averaged radiative loss profile in our simulation is consistent with that in the classic atmospheric model. In addition, the energy flux in the corona is larger than the required value for coronal heating in the quiet region. Our study shows that the Alfv\'en wave driven model has the potential to explain chromospheric heating and transport enough energy to the corona simultaneously.
\end{abstract}


\section{Introduction
  \label{sec:introduction:background}}

The problem how to heat the solar chromosphere and the corona to maintain their temperature at $10^4$ K to $10^6$ K is still under debate. On average, energy fluxes of $3 \times 10^5$ erg s$^{-1}$ cm$^{-2}$  and $2 \times 10^6$ erg s$^{-1}$ cm$^{-2}$ in the quiet region are required for coronal and chromospheric heating, respectively \citep{1977ARA&A..15..363W}. The chromosphere is divided into high-beta non-magnetic region below the magnetic canopy \citep{1976RSPTA.281..339G} and low-beta magnetic region in flux tubes and higher chromosphere above the equipartition layer ($\beta = 1$ layer). The height of the equipartition layer in the quiet region is 0.8 Mm - 1.6 Mm \citep{2014A&ARv..22...78W}. 

The dissipation of acoustic shock is considered to be a candidate for chromospheric heating in non-magnetic chromosphere \citep{1948ApJ...107....1S,1989ApJ...336.1089A,1993ApJ...414..337J}. The dynamics of acoustic wave propagation in non-magnetic chromosphere have been well studied by hydrodynamic simulation with non-local thermodynamic equilibrium (non-LTE) radiative transfer \citep{1995ApJ...440L..29C,1997ApJ...481..500C}, which show that the synthesized emerging Ca II K line spectra is consistent with the observation. 

However, it is difficult for acoustic waves to supply energy in high chromosphere \citep{1982A&A...106....9U,1993ApJ...414..337J}, as they dissapate energy fast at a lower position. At heights above the equipartition layer, low-beta magnetic regions occupy all the space where Alfv\'en wave \citep{1942Natur.150..405A} is considered as an important energy transporter \citep[e.g.][]{1947MNRAS.107..211A,2013SSRv..175....1M,2019ApJ...871....3S}. Numerical studies suggest that continuous input of transverse perturbations at the photosphere, which behave as Alfv\'en waves, could contribute to coronal heating (\citealt{1999ApJ...514..493K}, hereinafter KS99; \citealt{2010ApJ...712..494A}; \citealt{2010ApJ...710.1857M}, hereinafter MS10). At the same time, as the nonlinearity increases with expansion of the flux tube, the non-linear mode coupling \citep{1982SoPh...75...35H,1991A&A...241..625U,2003ApJ...587..806M} generates acoustic (slow mode) waves, which steepen to produce shocks and dissipate to provide energy for chromospheric heating \citep{2016ApJ...829...80B,2016ApJ...817...94A,2012ApJ...749....8M}. As a result, the scenario in which Alfv\'en waves carry energy to the higher chromosphere and the corona while the chromospheric heating is powered by the shock dissipation of longitudinal waves, which are initialized by the mode coupling from these Alfv\'en waves, has been promoted. 

Previous studies on Alfv\'en wave propagation in magnetic chromosphere have usually ignored or crudely treated the radiative loss in the chromosphere, which is the most significant source of energy loss \citep{1977ARA&A..15..363W}. MS10 and \cite{2012ApJ...749....8M} do include radiative loss while applying the approximation in \cite{1989ApJ...336.1089A}, where the radiative loss is only determined by the local density, which means that the chromospheric plasma has a constant cooling time. \cite{2016ApJ...829...80B} also include radiative loss, where the radiative loss rate at a certain position is determined by the time average of the viscous heating during the previous 160 s. However, as pointed out by \cite{1995A&A...293..166H}, radiative loss is much more narrowly concentrated in the hot region behind the shock. which cannot be correctly reflected by the treatments used in these studies. On the other hand, models with advanced 3D radiative MHD simulation \citep[e.g.][]{2017ApJ...848...38I,2011A&A...531A.154G} as well as synthesized observation is widely used in diagnostic of spectral lines formed in the chromosphere and the transition region \citep[e.g.][]{2013ApJ...772...90L,2015ApJ...811...81R}, however, their complexity adds the difficulty in understanding the underlying physical process.

To investigate the applicability of previous Alfv\'en wave driving model on chromospheric heating, we conduct MHD simulations with an improved treatment of radiative loss introduced by \cite{2012A&A...539A..39C} (hereinafter CL12). We ignore the longitudinal wave input at the photosphere to avoid mixture of mode-coupling initiated and the input longitudinal waves in the chromosphere. In this paper, we consider a similar geometry setting following KS99 and MS10. We study chromospheric heating by comparing the spatial distributions of the radiative loss profile in our simulation and the classic model VALC \citep{1981ApJS...45..635V}. The setting of our simulation is introduced in Section 2. The results are shown in Section 3. Discussions and comparison with previous study are included in Section 4. Finally, we summarize our results in Section 5.

\section{Numerical setting}\label{sec:2}

We solve 1.5D ideal compressible MHD equations on an expanding flux tube whose cross section area $A$ is a function of the height $z$, which does not change with time. The expression 1.5D indicates that we have a one-dimensional geometry setting while the velocity and magnetic field has two components, namely the $s$ direction and the $\phi$ direction. The $s$ direction is curved along the flux tube, while the $\phi$ direction is the azimuthal direction. The basic equations are  
\newcommand\abs[1]{\left|#1\right|}
 \begin{equation} 
\frac{\partial}{\partial t}(\rho A) + \frac{\partial}{\partial s}(\rho V_s A) = 0,
 \end{equation}
 \begin{equation} \label{eq:mhd2}
  \begin{aligned}
&\frac{\partial}{\partial t}(\rho V_s A) + \frac{\partial}{\partial s}[(\rho V_s^2 + P + \frac{B_\phi^2}{8\pi})A] \\
&= (P + \frac{\rho V_\phi^2}{2})\frac{dA}{ds} - \rho g_0 A \frac{dz}{ds},
\end{aligned}
\end{equation}
 \begin{equation} \label{eq:mhd3}
\frac{\partial}{\partial t}(\rho V_{\phi} A^{\frac{3}{2}}) + \frac{\partial}{\partial s}[(\rho V_\phi V_s - \frac{B_\phi B_s}{4\pi}]A^{\frac{3}{2}})] = A \rho L_{\text{trq}},
\end{equation}
 \begin{equation} 
\frac{\partial}{\partial t}(\sqrt{A} B_{\phi})+ \frac{\partial}{\partial s} [(B_\phi V_s - B_s V_\phi)\sqrt{A}] = 0,
\end{equation}
 \begin{equation} 
 \begin{aligned}
&\frac{\partial}{\partial t}[(\frac{\rho {\mathbf{V}}^2}{2} + \frac{P}{\gamma-1}+\frac{{\mathbf{B}}^2}{8\pi})A] \\&+ 
\frac{\partial}{\partial s}\{[(\frac{\rho {\mathbf{V}}^2}{2} + \frac{\gamma P}{\gamma-1} + \frac{B_\phi^2}{4\pi})V_s-\frac{B_s B_\phi V_\phi}{4\pi}]A\}\\ 
&= -L_{\text{rad}} A  - \rho V_s g_0 \frac{dz}{ds} A + \rho V_\phi\sqrt{A} L_{\text{trq}} + S_{\text{art}} A,
\end{aligned}
\label{eq:rad}
\end{equation}
and the ideal gas equation of state which is given by
 \begin{equation} \label{eq:ideal}
P=\frac{k_B}{m}\rho T,
\end{equation}
where $\rho$ is the density; $A$ is the cross-section area; $t$ is the time; $s$ is the distance along the field line; $V_s$ is the velocity along the $s$ direction; $P$ is the gas pressure; $V_\phi$ is the velocity along the $\phi$ direction; $g_0$ is the gravity; $z$ is the height; $B_s$ is the magnetic field along the $s$ direction, $B_s$ does not change with time and we set $B_s A$ to be a constant to obtain the divergence free condition of the magnetic field; $L_{\text{trq}}$ is the transverse torque; $B_\phi$ is the magnetic field along the $\phi$ direction; $\mathbf{V}$ is the velocity vector; $\mathbf{B}$ is the magnetic field vector; $L_{\text{rad}}$ is the radiative loss; $\gamma$ is the ratio of specific heats, $\gamma=\frac{5}{3}$; $T$ is the temperature; $m$ is the mass per particle, assuming $m= m_{\text{H}} = 1.67\times10^{-24}$ g; $k_B$ is the Boltzmann constant, $k_B=1.38\times10^{-16} \text{erg } \text{K}^{-1}$; and $S_{\text{art}}$ is an artificial heating term that is used to prevent the temperature from dropping too low. For the derivation of 1.5D MHD equations in curvilinear coordinates, one could refer to \cite{2018ApJ...854....9S}.

We set the expanding flux tube geometry following KS99 by setting the radius of the flux tube $r$ as a function of $z$. The radius $r$ and cross section area $A$ have the relation $A = \pi r^2$. The radius is given by

 \begin{equation}\label{eq:rrs0}
r= \int \cos \alpha ds ,
 \end{equation}
 \begin{equation}\label{eq:rrs1}
z= \int \sin \alpha ds ,
 \end{equation}
 where
  \begin{equation}\label{eq:rrs2}
\alpha= \alpha_{t} + (\alpha_r - \alpha_{t}) f_n ,
 \end{equation}
   \begin{equation}\label{eq:rrs3}
\alpha_r= -\text{arctan}(\frac{-4H_0}{r}) ,
 \end{equation}
 \begin{equation}\label{eq:rrs}
\alpha_{t}=\text{arctan}[k \text{ cosh}(z/z_d)^2].
 \end{equation}
    \begin{equation}\label{eq:rrs5}
f_n=-\frac{1}{2} \{\text{tanh} [(z-0.2z_d)/(0.1 z_d)]-1\}.
 \end{equation}

 We set $z_d=2250$ km and $H_0=150$ km following KS99. Also, $r$ as a function of $z$ is obtained by numerically solving the ordinary differential equation
 \begin{equation}\label{eq:rrsb}
\frac{\text{d} r}{\text{d} z} = \cot{\alpha}.
 \end{equation}
 
\begin{figure}[!h]
    \centering
    \includegraphics[width=8cm]{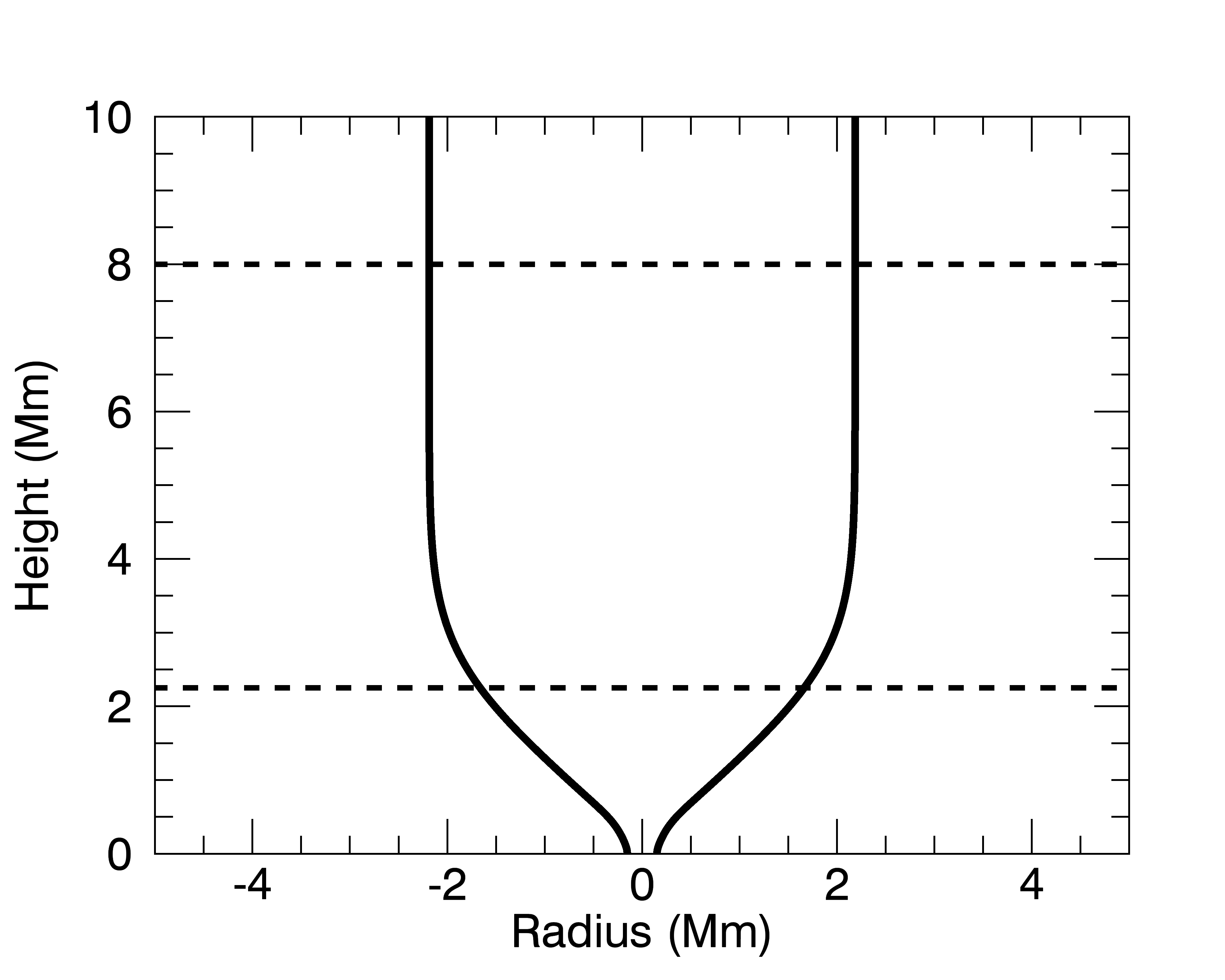}
    \caption{Longitudinal section of the expanding flux tube. The lower dashed line shows the height of the transition region, which is $2.25$ Mm in this study. The higher dashed line shows the height where the energy flux in the corona is measured, which is $8$ Mm.}
    \label{fig:2-1}
\end{figure}

The degree of expansion can be varied by adjusting $k$ in Equation (\ref{eq:rrs}). We set $k=1.2$ for a typical case, while $k$ is also adjusted for a parameter survey of the flux tubes with expansion factor changing within observational range. In addition, we keep the radius of the flux tube $r$ constant when $z < 0.1$ Mm and apply a Gaussian kernel smoothing with a width of 0.04 Mm to connect the low constant radius part and the expanding part of the flux tube. This setting mimics a flux tube expanding from the network region. The longitudinal section of the flux tube is shown in Figure \ref{fig:2-1}. The radius of the flux tube $r$ at the lower boundary $z=0$ is set to be $150$ km, which is approximately the length of the pressure gradient height. The starting point $s=0$ is at the same position where $z=0$. The expansion factor $f$ describes the degree of expansion of the flux tube, which is defined as
  \begin{equation}\label{eq:rs}
f=A_{\text{low}}/A_{\text{top}},
 \end{equation}
where $A_{\text{low}}$ and $A_{\text{top}}$ are the cross-section areas of the flux tube at the top boundary and lower boundary, respectively.

There are three extra terms beside an ideal MHD model. The first one is gravity. In our model, gravity $g_0$ is calculated by
 \begin{equation} 
g_0=\frac{GM_{{\sun}}}{(z+R_{{\sun}})^2},
\end{equation}
where $G$ is the gravitational constant, $M_{{\sun}}$ is the mass of the sun, and $R_{ {\sun}}$ is the radius of the sun.

\begin{figure}[!h]
    \centering
    \includegraphics[width=8cm]{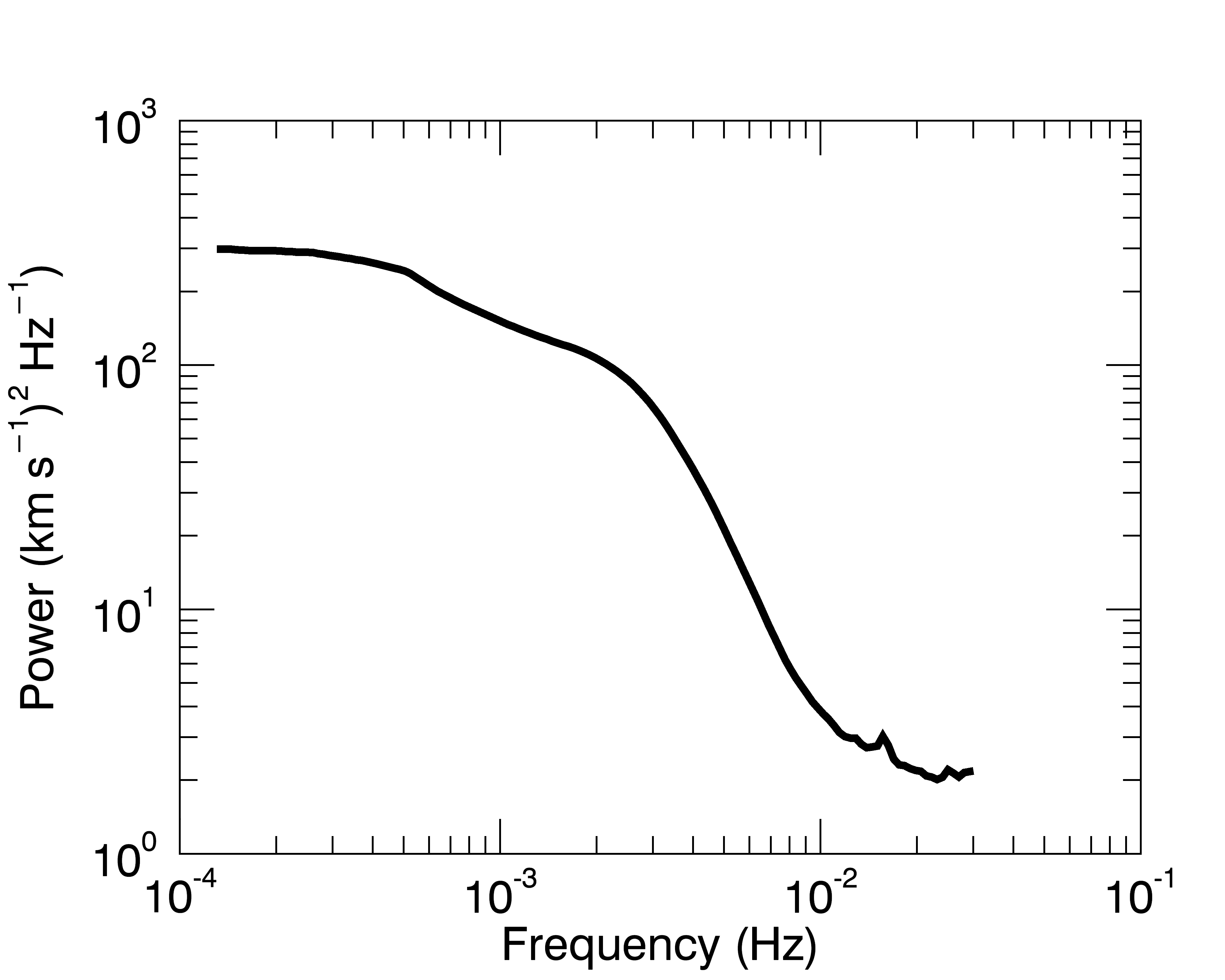}
    \caption{Observed power spectrum of photospheric horizontal velocity, modified from Figure 2 in MS10.}
    \label{fig:3-6}
\end{figure}

The second term is the transverse torque $L_{\text{trq}}$. The Alfv\'en wave is initialized by this transverse torque, which mimics the convection motion at the photosphere. Following KS99 and MS10, $L_{\text{trq}}$ is modeled to have the following form:
\begin{equation}\label{ltrq}
L_{\text{trq}}(t,z)=r W_0(t)(\tanh(\frac{z-0.75H_0}{0.075H_0})-1),
\end{equation}
where $W_0(t)$ determines the amplitude and time evolution of the artificial torque. We adjust the form and amplitude of the artificial torque by adjusting $W_0(t)$, which is derived from the velocity spectra. The transverse velocity at the bottom has the form
 \begin{equation}
V_{\phi}(t,z=0)= \sum_i C_i \sin(2\pi \nu_i t + \psi_i),
\end{equation}
where $C_i$ determines the power of the transverse velocity at frequency $\nu_i$ by using the veolicity spectra. Frequency $\nu_i$ is chosen to be 100 points averagely distributed between the chosen minimum frequecny $f_{\min}=2 \times 10^{-4}$ s$^{-1}$ and the maximum frequency $f_{\max}=5 \times 10^{-2}$ s$^{-1}$. $C_i$ is obtained from the spectra of the observed transverse velocity of the photosphere shown in Figure \ref{fig:3-6} (modified from Figure 2 in MS10). The phases $\psi_i$ are random numbers between $0$ and $2 \pi$ for each $i$. To obtain this velocity distribution, we set the intensity of torque to be the acceleration that has the form 

\begin{equation}\label{eqqq}
W_0(t) = \frac{dV_{\phi}}{dt}=  \sum_i 2\pi  \nu_i C_i \cos(2\pi \nu_i t + \psi_i).
\end{equation}

We apply a multiplier to $W_0(t)$ in order to adjust the root mean square of the transverse velocity at the lower boundary to be around $1$ km/s.

The third term is radiative loss $L_{\text{rad}}$. Following CL12, we have
 \begin{equation} \label{eq:r}
L_{\text{rad}}=-\sum_{X_m}L_{X_m}(T) E_{X_m}\frac{N_{X_m}}{N_{X}}(T)A_{X} N_{\text{H}} n_e.
\end{equation}

Here, the subscript $X_m$ represents a component of element $X$ in the ionization state $m$. $X_m$ used in this approximation method include neutral hydrogen (H I), singly ionized calcium (Ca II), and singly ionized magnesium (Mg II), since they are the most important components for chromospheric radiative loss \citep{1981ApJS...45..635V}. $A_X$ is the abundance of element $X$. $L_{X_m}(T)$ represents the optically thin radiative loss for different elements that are functions of $T$. $\frac{N_{X_m}}{N_{X}}(T)$ represents the fractions of specific ions or neutral atmos in the ionization state $m$ of element $X$, which are functions of $T$. $E_{X_m}$ is the escape probability. \rv{Escape probability are tabulated functions of column mass for Mg II and Ca II and neutral hydrogen column denstiy for H. One could refer to Section 4.2 in CL12 for further explanation}. The column mass is calculated by \(\int_z \rho dz\). The neutral hydrogen column density is calculated by \(\int_z \rho/m_{\text{H}}\frac{N_{\text{H I}}}{N_{\text{H}}}(T)\). All these functions are obtained by fitting with a detailed radiative transfer calculation. $N_{\text{H}}$ is the number density of hydrogen element and $n_e$ is the number density of electrons; $N_{\text{H}}$ is determined by substituting temperature into the function of the fraction of neutral hydrogen. We assume $n_e = \rho/m_{\text{H}} - N_{\text{H}}$. The aim of this approximation approach is to obtain a simple form of $L_{X_m}(T)$, $E_{X_m}$, and $\frac{N_{X_m}}{N_{X}}(T)$ as a function of some physical parameters, so that we can calculate the radiative loss rate by putting proper values into these functions without carrying out complete radiative transfer calculations.

Heat conduction is not included in the simulation since the timescale for heat conduction in the chromosphere is much longer than the wave transition time. In addition, we also ignore the radiative loss in the corona since we mainly focus on the chromosphere and we have a very crude grid size in the corona. As we also ignore heat conduction, we cannot treat the energy balance in the corona carefully.

For the initial condition, we assume a hydrostatic stratified atmosphere in which
   \begin{equation}\label{eq:i1}
 \frac{d P}{d z}=- \rho g_0.
\end{equation}

\begin{figure}[!h]
    \centering
    \includegraphics[width=8cm]{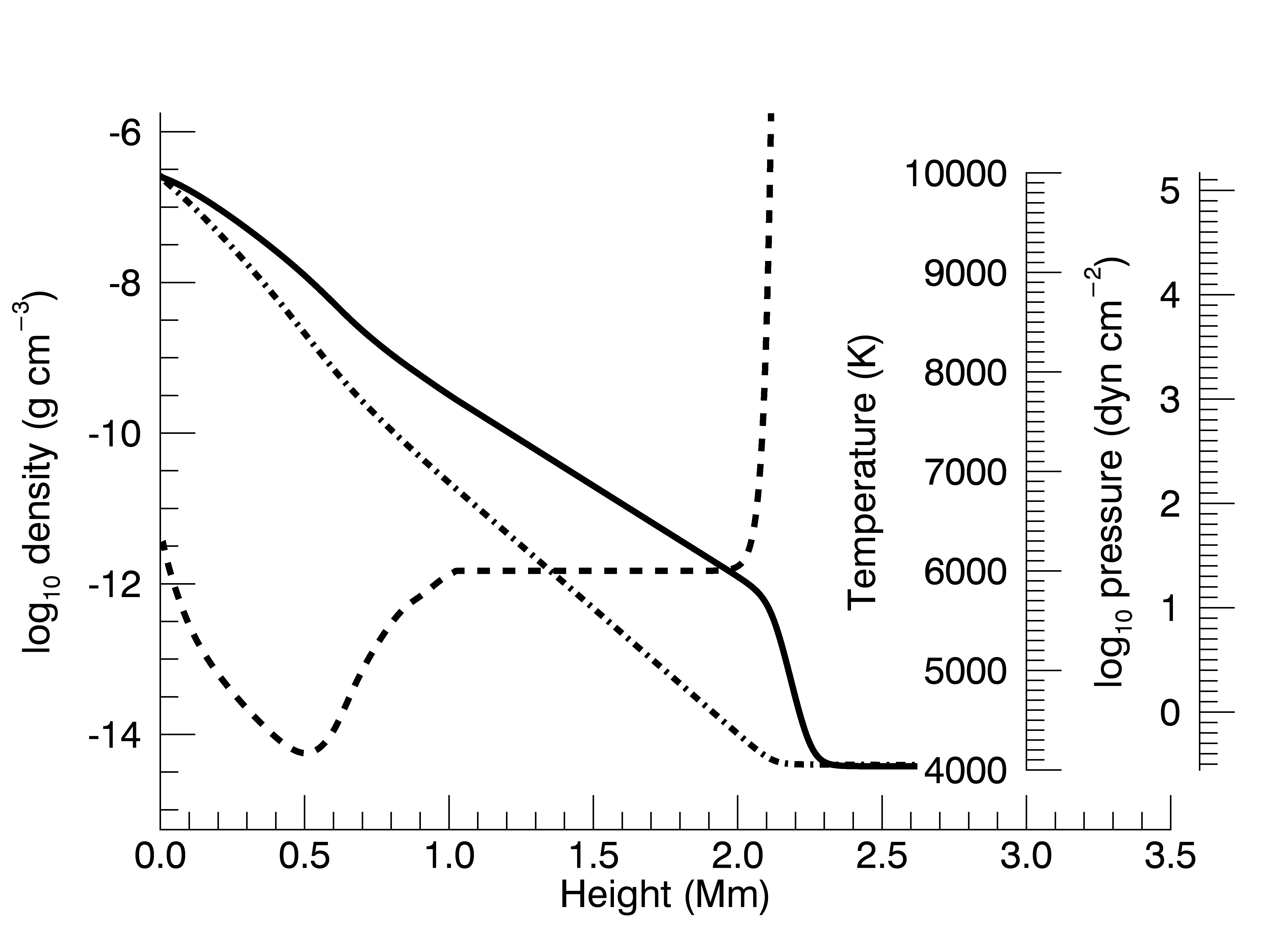}
    \caption{Temperature, density, and pressure of initial atmosphere as a function of height $z$, where the solid, dashed, and dash-dotted line represent the density, temperature, and pressure, respectively.}
    \label{fig:init}
\end{figure}

\begin{figure}[!h]
    \centering
    \includegraphics[width=8cm]{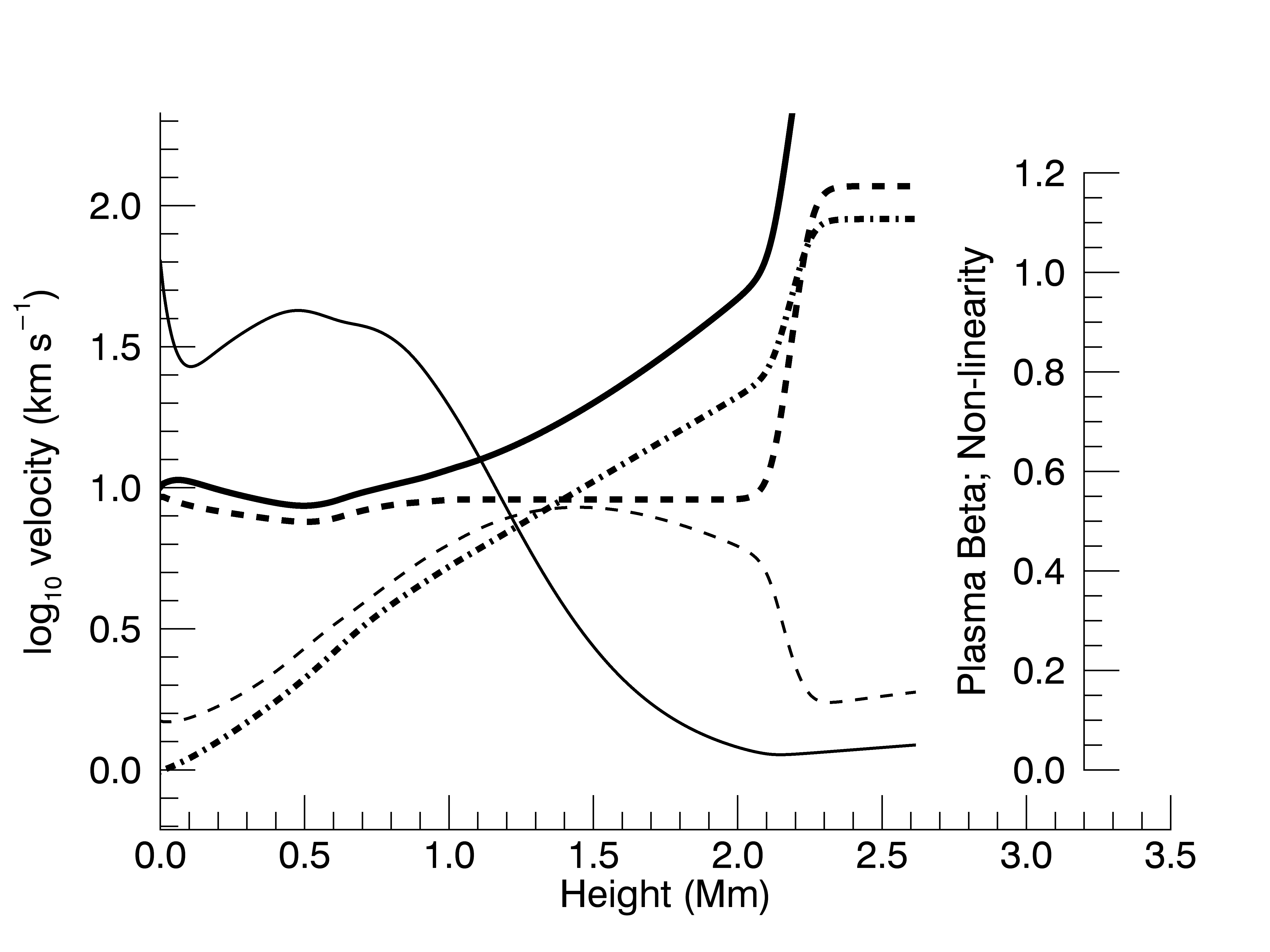}
    \caption{Alfv\'en speed (thick solid), sound speed (thick dashed), $v_{\phi\text{WKB}}$ (thick dash-dotted), plasma beta (thin solid), and non-linearity of Alfv\'en wave (thin dashed) as functions of height $z$.}
    \label{fig:init2}
\end{figure}

The initial temperature distribution is a combination of the classic VALC temperature model and a hyperbolic tangent distribution that is described below 
     \begin{equation} \label{eq:i2}
     T=
 \begin{cases} 
      T_{\mathrm{valc}}(z) & z \le 1\text{ Mm}\\
      T_{\mathrm{pho}}+\frac{1}{2}(T_{\mathrm{cor}}-T_{\mathrm{pho}})(\tanh(\frac{z-z_{\text{tr}}}{w_{\text{tr}}})+1) &  z > 1\text{ Mm}
   \end{cases},
\end{equation}
where $T_{\mathrm{valc}}(z)$ is the temperature distribution as a function of height in the VALC model; $T_{\text{cor}}$ is the temperature of the corona, which is set to be $10^6 $ K; $T_{\text{pho}}$ is the temperature of the photosphere, which is set to be $6000$ K; $z_{\text{tr}}$ is the height of the transition region, which is set to be $2.25 $ Mm; $w_{\text{tr}}$ relates with the width of the transition region, which is set to be $0.05$ Mm. The density at lower boundary is set to be $2.53 \times 10^{-7} \text{ g cm}^{-3}$. After the temperature is determined, the pressure and density are calculated as functions of height by using Equation (\ref{eq:i1}) and the equation of state of ideal gas (Equation \ref{eq:ideal}). The distributions of temperature, gas pressure, and density are shown in Figure \ref{fig:init}. The background Alfv\'en speed, sound speed, plasma beta, and non-linearity of the Alfv\'en wave are shown in Figure \ref{fig:init2}. The non-lineartiy of the Alfv\'en wave is estimated by $v_{\phi\text{WKB}}/C_A$, where $C_A$ is the background Alfv\'en speed and $v_{\phi\text{WKB}}$ is the amplitude of the wave in the azimuthal direction estimated by WKB approximation. $B_s$ at the photosphere is determined by the gas pressure required to maintain the plasma beta around unity. As a result, the magentic field at the bottom is 1812 G. The pressure at the bottom is $1.26 \times 10^5 \text{ dyn cm}^{-2}$.

The MHD equations are solved using the upwind scheme with Modified Harten-Lax-van Lee approximate Riemann solver \citep[HLLD;][]{2005JCoPh.208..315M}. We set the scheme to have second-order accuracy in terms of space and time by applying the Monotonic Upwind Scheme for Conservation Laws (MUSCL) reconstruction \citep{1979JCoPh..32..101V} with minmod slope limiter \citep{1986AnRFM..18..337R} and the second order Total Variation Diminishing Runge-Kutta scheme \citep{1988JCoPh..77..439S} for time evolution. At the lower boundary, the density and pressure of the point at the outer boundary increase according to the hydrostatic stratification. For the momentum perpendicular to the boundary and $B_{\phi}$, it has the same absolute value but opposite directions. The other physical parameters parallel to the boundary are symmetric. The top boundary is a free boundary. \rv{There is reflection of waves at the top boudnary, it is more ideal if we could have an open boundary for waves propagating freely across the top boundary, however, since Alfv\'en wave are highly reflected at the transition region \citep{2005ApJS..156..265C}, the energy flux of the Alfv\'en wave in the corona is too small to affect the chromosphere. As a result, we can ignore the reflected wave from the top boundary.} We simulate up to 9 Mm with an evenly distributed grid having a size of around 5 km. Above 9 Mm, the length of each grid increases gradually. The value of $z$ at the top of the simulation region is 200 Mm.

\section{Results} \label{sec:3}
\begin{figure}%
\centering
\includegraphics[width=8cm]{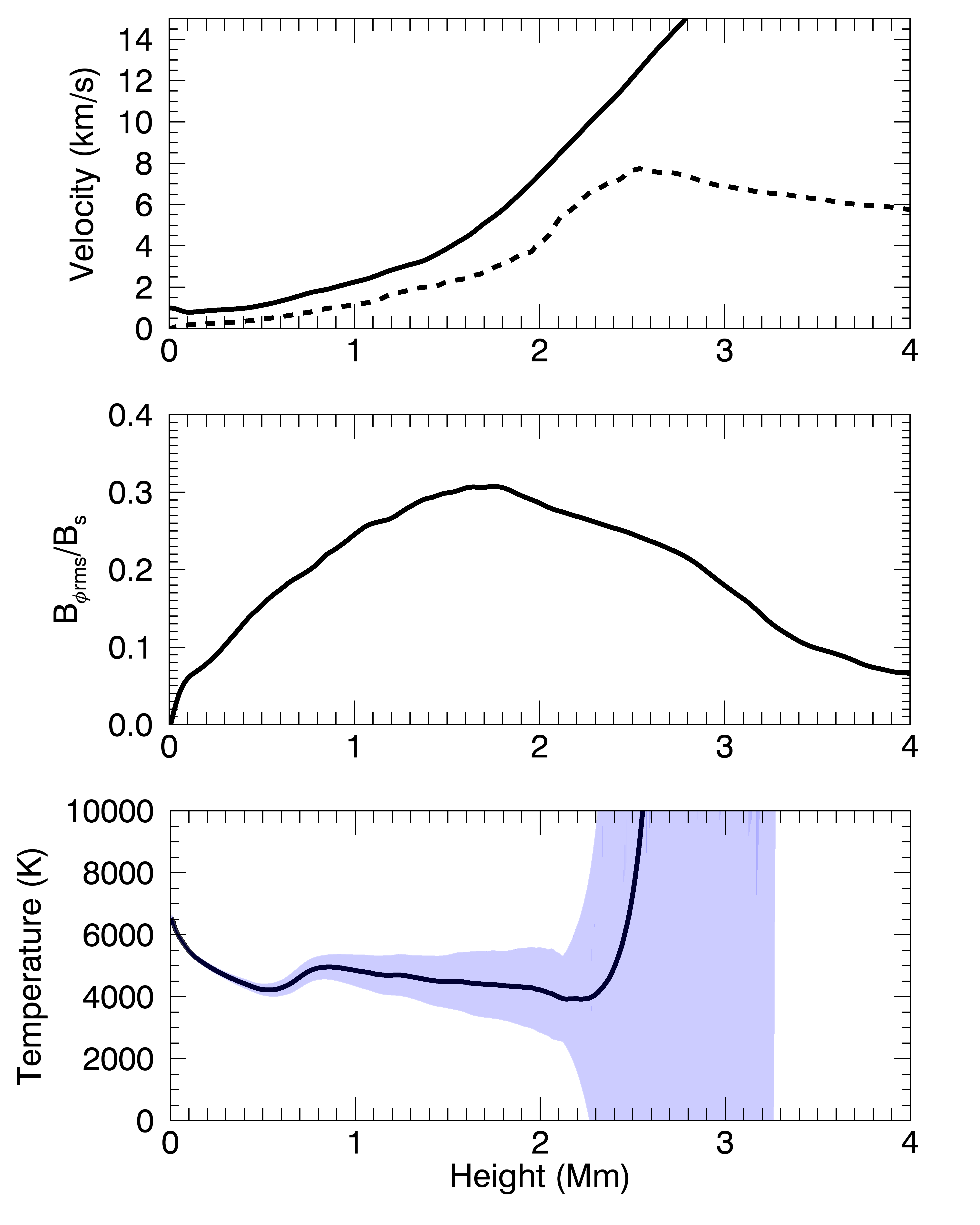}
\caption{Upper panel: root mean square of transverse (solid) velocity and longitudinal (dashed) velocity over time. Middle panel: root mean square of transvere magnetic field over time, normalized by longitudinal magnetic field, which does not change with time. Lower panel: time-averaged temperature. The blue region marks the region with plus and minus one standard deviation.}
   \label{fig:ta}
\end{figure}

\begin{figure}%
\centering
\includegraphics[width=8cm]{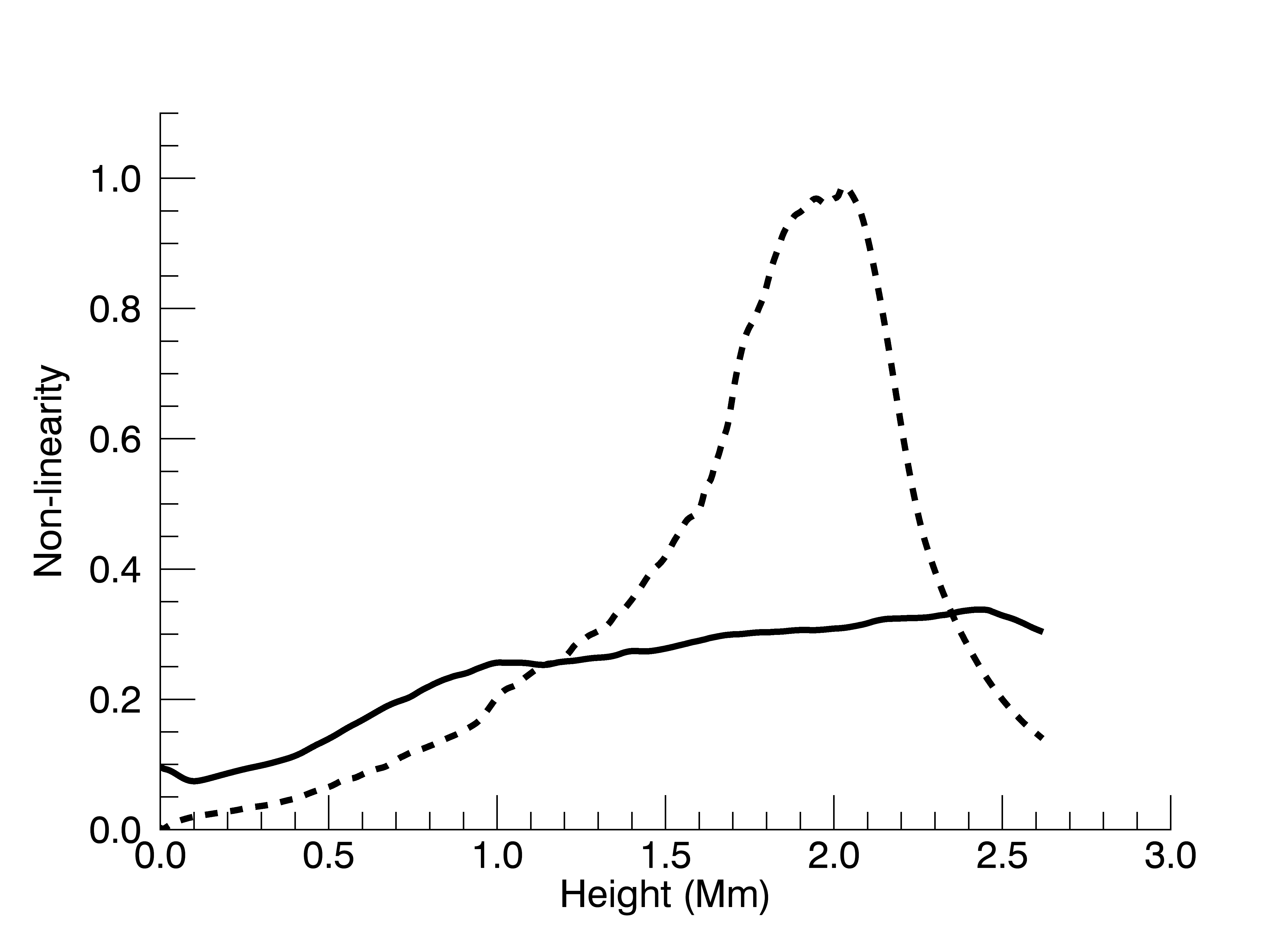}
\caption{Non-linearity of transverse (solid) and longitudinal (dashed) wave for a typical case.}
   \label{fig:nl}
\end{figure}

The root mean square of the velocity and transverse magnetic field over time as well as the time-averaged temperature in the chromosphere for a typical case are shown in Figure \ref{fig:ta}. The waves in the chromosphere are shown by the non-linearity of the time-averaged velocity, which is defined by the root mean square of the transverse (longitudinal) velocity divided by the time-averaged Alfv\'en (sound) speed (Figure \ref{fig:nl}). An increse in non-linearity with height indicates steepening of waves as they propagate upwards, especially for longitudinal waves.

An ideal way of making a comparison with the observation is synthesizing the emerging spectra and comparing them with the observation. However, it is difficult to perform synthesis in the chromosphere due to the NLTE condition in the chromosphere and the limitation of 1.5D geometry \citep{2017A&A...597A..46S}. Instead, we compare the radiative loss profile and temperature in our simulation with those of the classic model. Since the time scale of radiative loss is around 200 s in the chromosphere and our calculation lasts 5000 s, which is around several tens of times that of the radiative cooling time. \rv{We expect that statistically, energy balance between heating and radiative loss in the chromosphere has already been reached that time-averaged cooling rate is identical to time-averaged heating rate. We will give further estimation of heating rate in the discussion part.}

\begin{figure}%
\centering
\includegraphics[width=8cm]{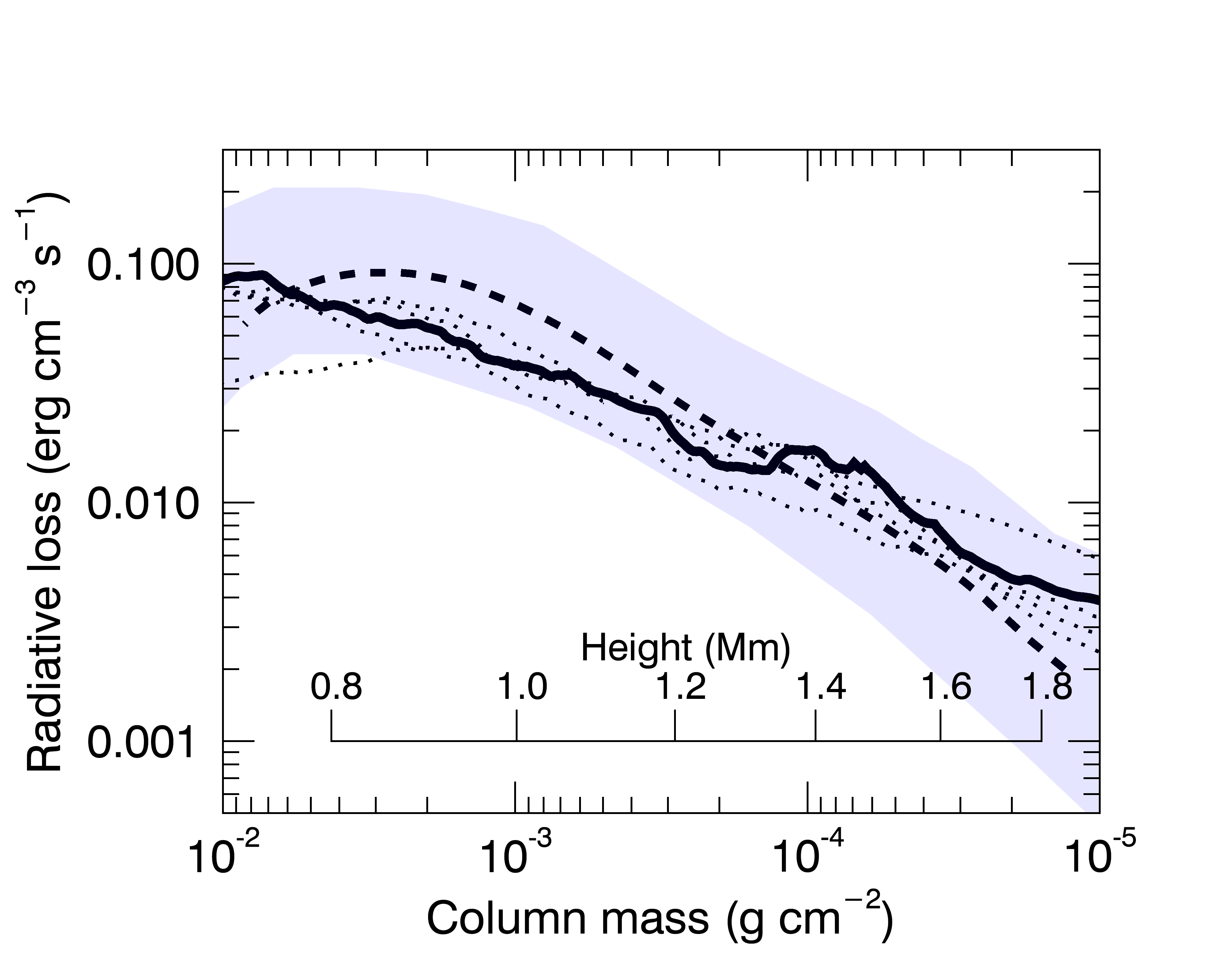}
\caption{Effective radiative loss rate as a function of column mass. The thick black solid line shows the radiative loss profile as a function of column mass for the typical case. The corresponding height in the VALC model is shown in the secondary axis. The dotted lines represent comparative simulations with adjustments in the background magnetic field. The VALC radiative loss profile is plotted by the thick dashed line and the blue region represents the radiative loss profile from Avrett Model A to Avrett Model F.}
   \label{fig:rad}
\end{figure}

The time-averaged effective radiative loss (ERL) profile for the typical case is shown by the thick black line in Figure \ref{fig:rad}. The effective radiative loss is defined as 
\begin{equation}\label{eq:eff}
L_{\text{ERL}}= L_{\text{rad}} A/A_c,
\end{equation}
where $A$ is the cross-section area at that height, $A_c$ is the cross-section area at the corona (defined at $z = 8$ Mm), and $L_{\text{ERL}}$ is the radiative loss rate with compensation of the expanding effect. Also, instead of being applicable just inside the flux tube, the effective radiative loss represents the averaged value across an entire slice of the cylinder, which has a constant cross-section A$_\text{c}$. \rv{We define the effective radiative loss since we are only focusing on the flux tube region and we want to emphasize that only the heating inside the flux tube could provide required heating for the chromosphere.} We plot the profile as a function of the column mass instead of height to prevent the influence from height variation of the transition region caused by formation of spicules. In addition, the radiative loss in the classic atmospheric model VALC is overplotted by the thick dashed line. The dotted lines represent the results of simulations with adjustments in the background magnetic field: change in the magnetic field at the bottom from 1 kG to 2 kG,
which is consistent with previous observation \citep{1989A&ARv...1....3S} and change in expansion factor from 0.003 to 0.015, within a reasonable range (0.002-0.02, \cite{1990A&A...234..519S}). The blue region represents the radiative loss profile between Avrett Model A and Avrett Model F \citep{1981spss.conf..173A}, where the radiative loss profile is given as a function of height. The conversion from height to column mass is based on Table 10 and Table 15 in \cite{1981ApJS...45..635V}. 

Our simulation results suggest that despite the change in the background magnetic field, the radiative loss profile in the simulation agrees quantitatively with the classic solar atmospheric model.
 
The energy flux in the corona (defined at $z = 8$ Mm) is $7.3 \times 10^{5}$ erg cm$^{-2}$ s$^{-1}$.  The energy flux required for coronal heating in the quiet region is $3.0 \times 10^{5}$ erg cm$^{-2}$ s$^{-1}$ \citep{1977ARA&A..15..363W}. In the calculations, with adjustment in magnetic field as described above, the largest and smallest fluxes are $2.21 \times 10^{6}$ erg cm$^{-2}$ s$^{-1}$ and $3.5 \times 10^{5}$ erg cm$^{-2}$ s$^{-1}$, respectively. We conclude that in these simulations, enough energy, which could meet the requirement of coronal heating in the quiet region, is transported to the corona. This result is consistent with KS99 and MS10. We also notice that the energy flux in the the typical case is much larger than that in MS10; we will discuss this in Section \ref{sec:4}. 

\begin{figure}[!h]
    \centering
    \includegraphics[width=8cm]{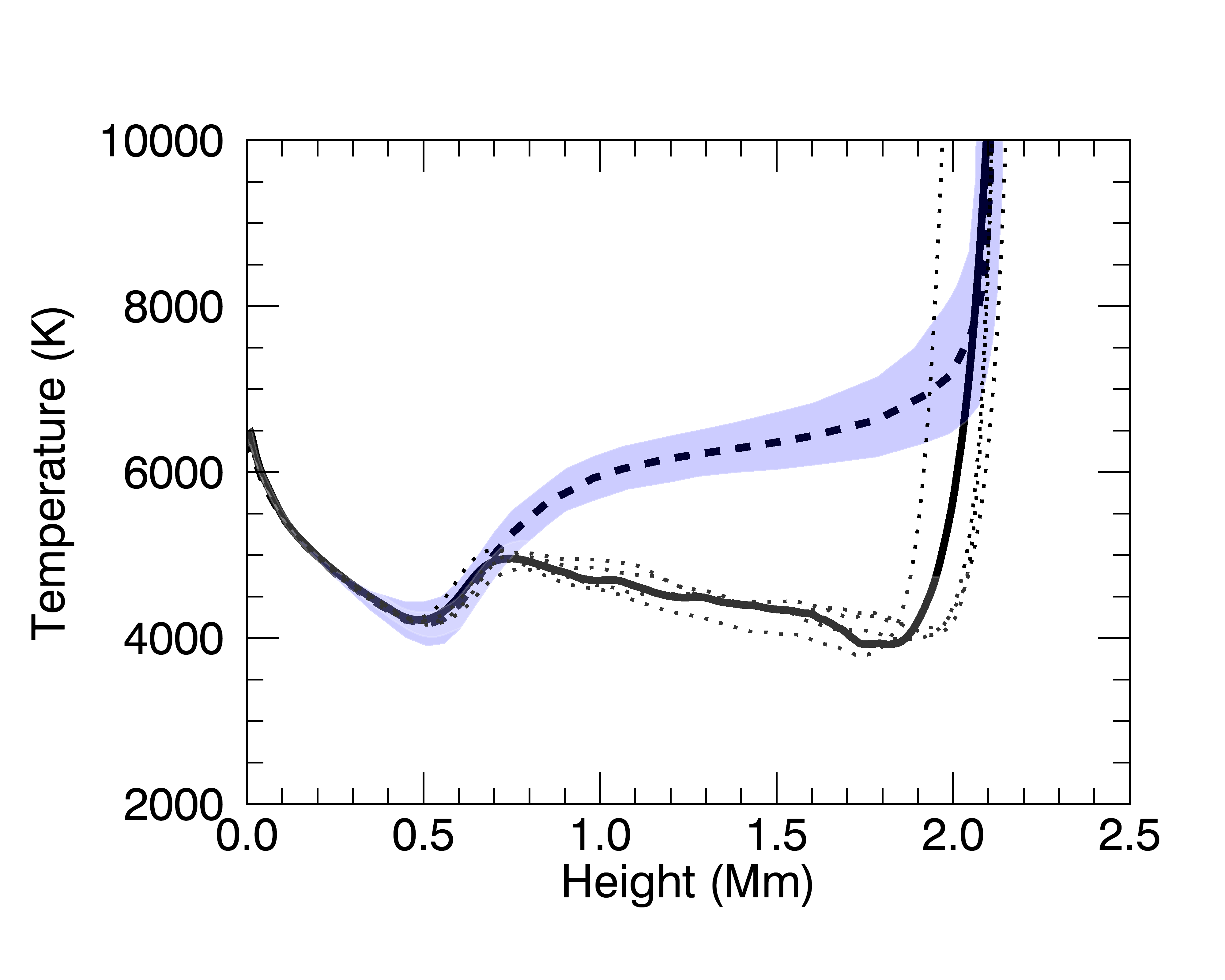}
    \caption{Time-averaged temperature as a function of height. The thick black solid line represents the typical case. The dotted lines are for comparative simulations with adjustments in the background magnetic field. The VALC temperature profile is plotted by the thick dashed line and the blue region represents the temperature profile from VALA to VALF model \citep{1981ApJS...45..635V}}
    \label{fig:4-1}
\end{figure}

The time-averaged temperature as a function of height is shown in Figure \ref{fig:4-1}, where the thick solid black line represents the time-averaged temperature profile and the thick dashed line represents the VALC temperature profile. Despite the result that the radiative loss profile is consistent with the classic atmospheric model, the time-averaged temperature profile is apparently lower than that in the classic model. In Figure \ref{fig:4-2}, from the upper panel to the lower panel, the time-integrated effective radiative loss, effective radiative loss, and temperature at a certain height $z=1.5$ Mm are shown in thick black solid lines. The slope of the dash-dotted lines in the upper panel represents the corresponding radiative loss rate at this height in the VALC model. We notice that when the shock front propagates across this height, as shown by the high temperature in the lower panel, a sudden increase in radiative loss occurs as shown in the comparison between the middle panel and lower panel. Also, in the upper panel, we notice that there are corresponding jumps, which indicate strong radiative loss. As a result, a continuous shock wave could support enough radiative loss. However, the low-temperature region between the shocks dominates most of the time, which leads to a lower time-averaged temperature. Low temperature without a temperature increase in the chromosphere is also obtained in other dynamic chromospheric models \citep{1994chdy.conf...47C,2004A&A...414.1121W}. \cite{1994chdy.conf...47C} suggest that \rv{the averaged gas temperature in the dynamic model is lower than that in hydrostatic equilibrium model despite that both of the two models have similiar emerging intensities. This is because high temperature shocks make a significant contribution to intensity in dynamic model. We need to point out that \cite{1994chdy.conf...47C} focuses on the non-magnetic region, which is different from our simulation, but the effect of shocks that cause the difference between the averaged gas temperature in the dynamic model and the hydrostatic equilibrium model is similiar.}

\section{Discussion
  \label{sec:4}}
\begin{figure}%
\centering
\includegraphics[width=8cm]{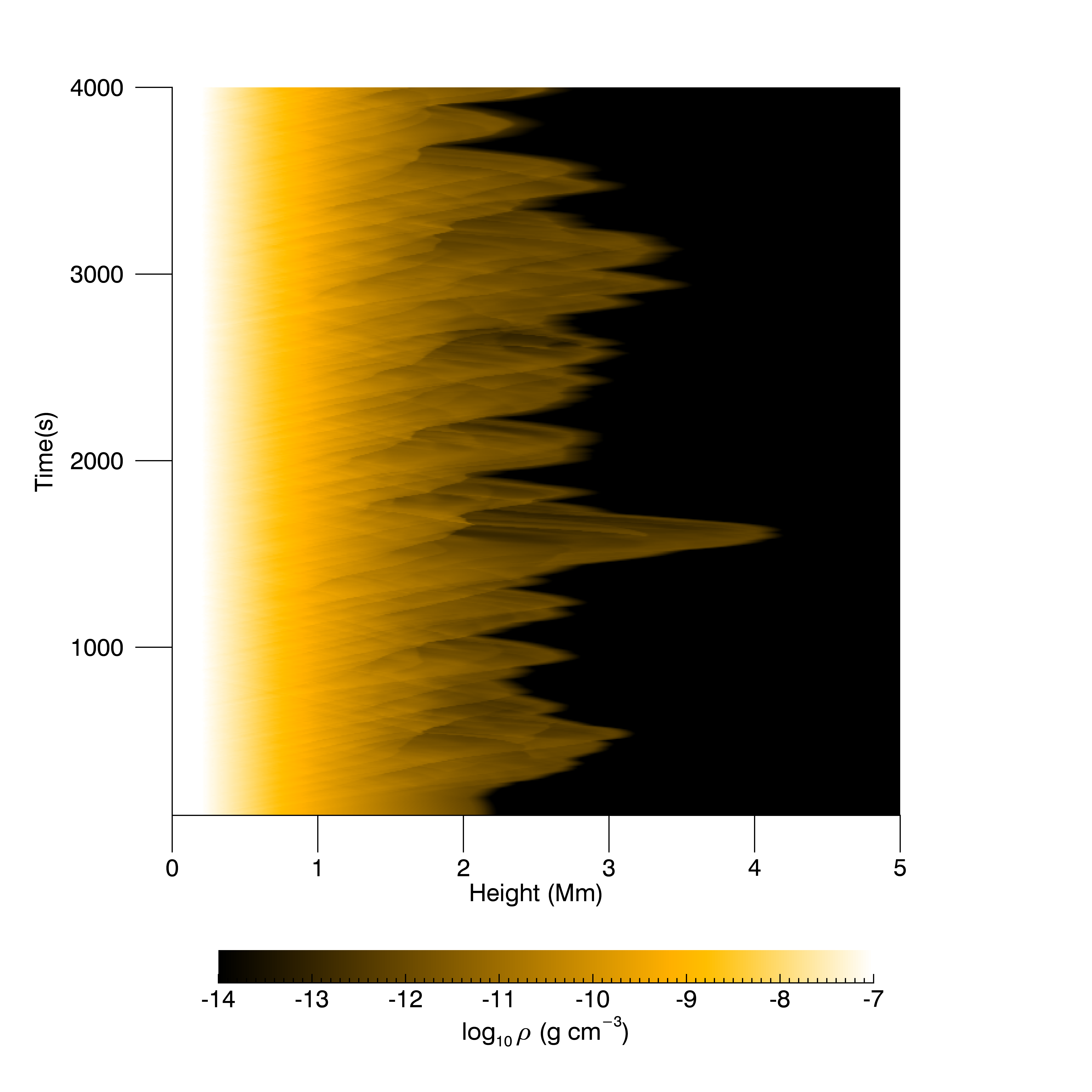}
\caption{Time slice of the distribution of density for the typical case.}
   \label{fig:ro}
\end{figure}

The time slice of the density distribution is shown in Figure \ref{fig:ro}. The rise and fall of the transition region reflect the formation of spicules. 

  \begin{figure*}
    \centering
    \includegraphics[width=\textwidth]{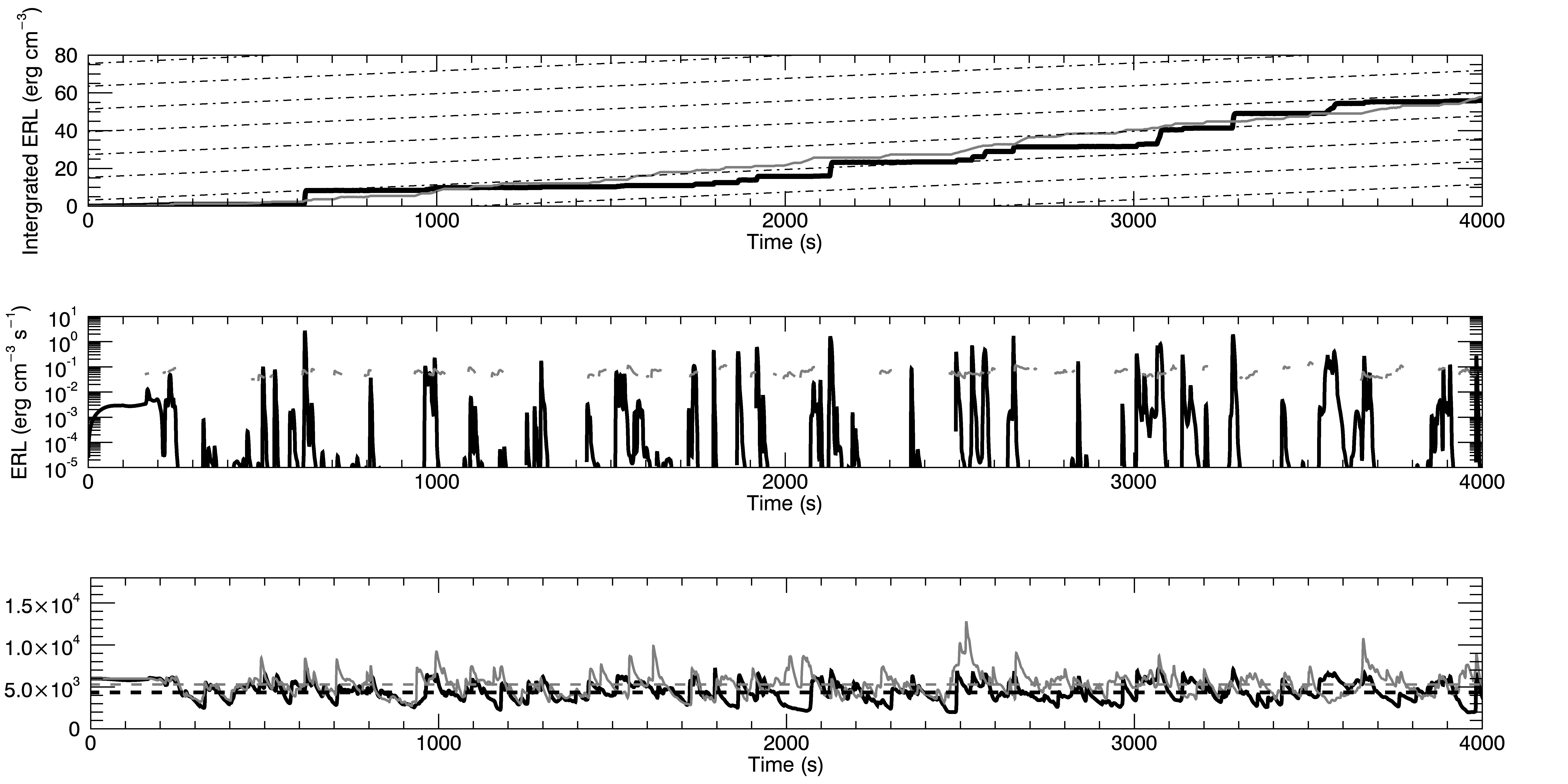}
    \caption{Upper panel: time integrated effective radiative loss. Middle panel: effective radiative loss. Lower panel: temperature. In all three panels, $x$ axis represents time. Height is fixed at 1.5 Mm. The black lines are for the typical case. The gray lines are for simulation with simplified radiative loss. The dash-dotted lines in the upper panel represent the corresponding radiative loss rate at 1.5 Mm in the VALC model. In the lower panel, black and gray dashed lines represent the time-averaged temperature at 1.5 Mm for the typical case and comparative simulation with simplified radiative loss, respectively.}
    \label{fig:4-2}
\end{figure*}

Our calculation is an extension of MS10, where they only apply crude treatment of radiative loss. In their study, by including only observation based transverse wave driver, enough energy is transported to the corona while achieving spicule formation at the same time. In our simulation, besides the energy flux in the corona and spicule formation, we emphasize that the time-averaged radiative loss profile is also consistent with the classic atmosphere model. Our result is consistent with \cite{2016ApJ...829...80B} in that we obtain a radiative loss profile that is consistent with the classic model. However, our treatment of chromospheric radiative loss is different. In addition, we also perform a parameter survey for making changes in the magnetic field intensity and expansion factor, which confirms the robustness of this result.

In our simulation, the energy flux in the corona is $7.3 \times 10^{5}$ erg cm$^{-2}$ s$^{-1}$, which is around 2 times of that in MS10; we conclude that this is mainly caused by the difference in stratification. In our simulation, the temperature in the \rv{photosphere} is 6000 K compared with 5000 K in MS10, and therefore we have a longer \rvv{scale height in the photosphere}, which leads to a higher density below the transition region. As a result, \rvv{the Alfv\'en} wave has a smaller \rv{phase} velocity in our calculation, and hence a shorter wavelength, \rv{Although a shorter wavelength will increase the dissipation rate of Alfv\'en wave in the chromosphere}, it also makes the \rv{transmittance} at the transition region become higher in our simulation. \rv{As a result, higher transmittance increases the energy flux in the corona.}   

\rv{We notice that the height of spicules is shorter \rvv{than} that in KS99 and MS10. For KS99, there is no radiative loss in the chromosphere and the internal energy in the chromosphere increase constantly. As a result, the height of spicules increases with time. For comparison with MS10, due to the difference in stratification, our simulation has a higher density below the transition region, which leads to a result that the spicule height in our simulation appears lower.}

\rv{We also estimate the heating rate and compare the time-averaged heating rate with radiative loss. 
Since our simulation does not contain explicit disspation, we estimate the heating rate at shock fronts from physicial parameters at both upstream and downstream region. The positions of shock fronts are identified by local minimum of $\frac{\partial {V_s}}{\partial s}$ with}
\begin{equation}\label{eq:sel}
\frac{\partial {V_s}}{\partial s} \le -\frac{1}{t_{\text{c}}}
\end{equation}
\rv{where $t_{\text{c}}$ is a parameter showing the threshold for shock wave identification. We chose $t_{\text{c}}$ changes between 10 s and 30 s. The selection of $t_{\text{c}}$ will be disscussed in detail in Appendix \ref{app1}. After identification of shock front, we choose local minimum or maximum of $\frac{\partial^2 {V_s}}{\partial s^2}$ near the shock front as the position to pick up upstream and downstream physical parameters. Heating rate is finally calculated following \cite{2007ApJS..171..520C} and specially averaged within the whole shock region, as shown below}
\begin{equation}\label{eq:cal}
Q_{\text{heat}}=c_v u_1 \rho_1 (T_2-T_1(\rho_2/\rho_1)^{\gamma})/w_{\text{shock}}
\end{equation}
\rv{where $Q_{\text{heat}}$ is heating rate per unit volume; $c_v$ is specific heat capacity at constant volume per unit mass; $T_1$ and $\rho_1$ are temperature and density at upstream region; $T_2$ and $\rho_2$ are those at downstream region. $u_1=v_1-u$ is velocity of upstream region in the shock rest frame, where $u$ is the propogating speed of shock front, $v_1$ is velocity at the upstream region. $u$ is calculated using the jump condition of conservation of mass. $u=(\rho_1 v_1-\rho_2 v_2)/(\rho_1-\rho_2)$. $w_{\text{shock}}$ is the width of shock wave, which is set to be 35 km. $w_{\text{shock}}$ dose not affect the total amount of heating rate (see Appendix \ref{app1}). The result is shown in Figure \ref{fig:heating}. We conclude that in the selected range of $t_{\text{c}}$, the estimated heating rate is found to be consistent with the radiative cooling rate. This result justify our usage of cooling rate as approximated value for heating rate.}

\begin{figure}%
\centering
\includegraphics[width=8cm]{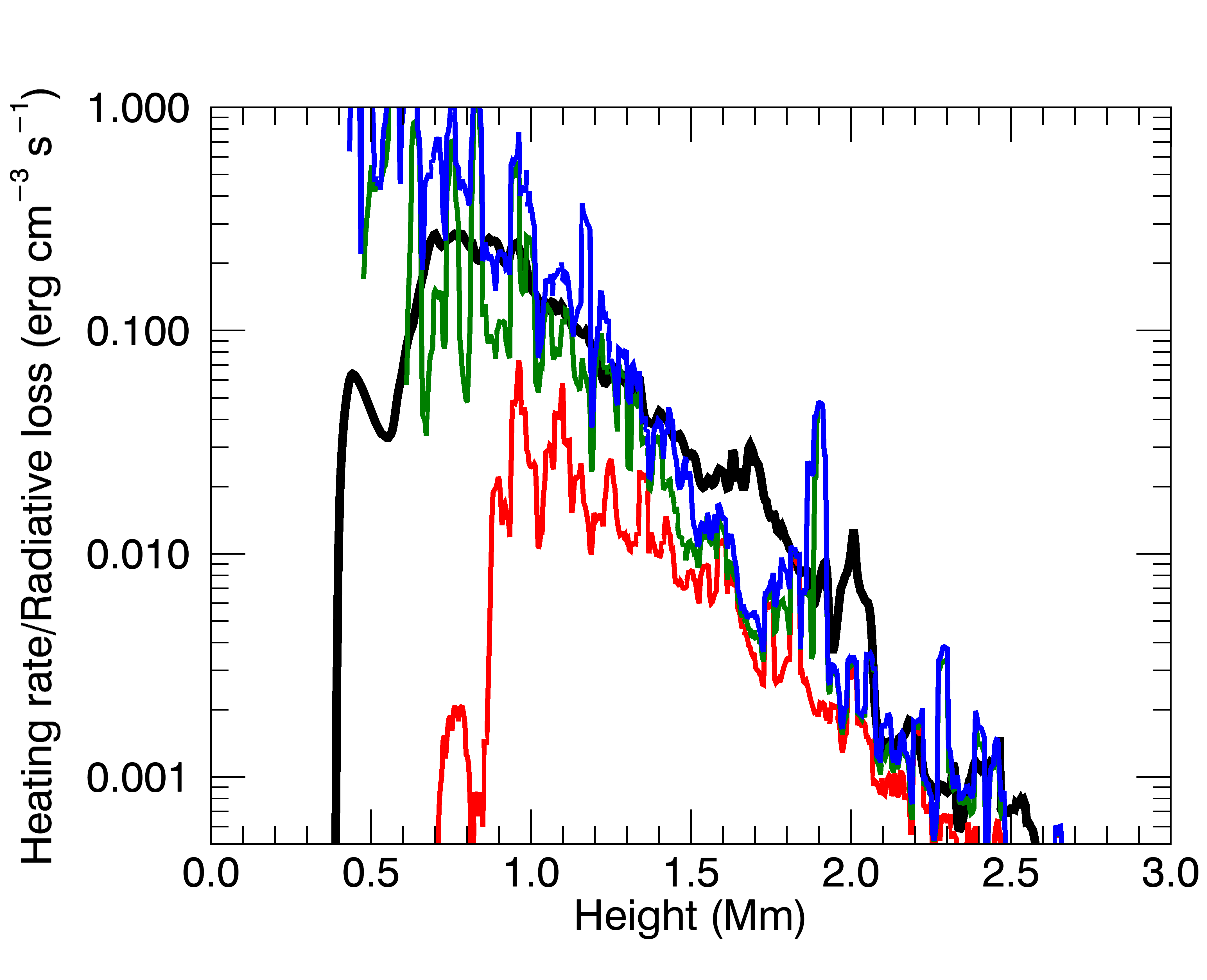}
\caption{Time-averaged radiative loss (thick black line) and estimated heating rate (thin colorful lines). Red, green and blue lines correspond to esimation of heating rate with $t_{\text{c}}=$10 s, 20 s and 30 s.}
   \label{fig:heating}
\end{figure}

\rv{For comparison with previous studies,} we also perform the simulation with only simplified radiative loss $L_{\text{rad}} = 4.9 \times 10^9$ erg cm$^{-2}$ s$^{-1}$. which is included in MS10. The result of temperature and radiative loss at $z=1.5$ Mm is shown in Figure \ref{fig:4-2}. From the upper to the lower panel, the time-integrated effective radiative loss, effective radiative loss, and temperature for the simulation with simplified radiative loss are shown by thin gray solid lines. This simplification method assumes a constant cooling time across the entire chromosphere. In our calculation, we set a lower limit T = 6000 K to switch on the simplified radiative loss. This lower limit is compulsory, since at the lower temperature region between shocks, the cooling time scale is much shorter than the acoustic wave transition time scale. If the simplified radiative loss is included without a switch, it will cause the temperature in regions between shock fronts to decrease to 0 when propagating upwards (or if artificial heating is included, fixed at the temperature below which artificial heating will take effect). We notice that the radiative loss at shock fronts in the simulation with simplified radiative loss is significantly smaller than that with CL12 radiative loss. According to the Rankine-Hugoniot condition, the compression ratio is smaller than 4, which results in a maximum of 4 times increase in the radiative loss at shock fronts compared with their surroundings. However, in detailed radiative transfer calculations, the radiative loss at shock fronts could be a few orders larger than that in the surroundings (e.g. Figure 14 in CL12, Figure 2 and Figure 3 in \cite{1995A&A...293..166H}). This leads to an underestimation of the radiative loss at shock fronts. \rv{We also estimate the robustness of simplified radiative loss term. We apply the simplified radiative loss term with different low temperature limits (6000 K and 5000 K). We find that the low temperature limit will directly affect the stratification in the chromosphere, which further leads to difference in height of the spicules and the energy flux of waves due to the reason that we have discussed above. This result suggests that there is a risk of loss of self-consistency when applying simplified radiative loss.} Therefore, we consider that although $L_{\text{rad}} = 4.9 \times 10^9$ erg cm$^{-2}$ s$^{-1}$ is a good approximation for time-averaged chromospheric radiative loss rate, one should be careful while applying this method to simulations studying chromospheric dynamics.

Our model is limited by the 1.5D geometry of a fixed flux tube. Besides, as we only focus on transverse wave, mode conversion \citep{2008SoPh..251..251C} from acoustic wave to Alfv\'en wave is ignored, although it is considered important for the generation of high-frequency Alfv\'en waves \citep{2018ApJ...854....9S}. In this simulation, we also ignore the longitudinal acoustic wave input at the photosphere to avoid mixture of mode-coupling initiated and input acoustic waves in the chromosphere. However, a comprehensive understanding of the role of waves in heating the magnetic chromosphere requires identifaction of different wave modes in the chromosphere and a thorough consideration of other heating mechanisms. A comparison between shock heating, turbulence heating \citep[][]{2011ApJ...736....3V}, ambipolar diffusion \citep[][]{2005A&A...442.1091L,2012ApJ...747...87K,2018A&A...618A..87K}, and other heating mechanisms is further desired.

\section{Conclusion
  \label{chap:summary}}

We solve 1.5D ideal MHD equations with CL12 approximated radiative loss model. We found that if observation based transverse perturbation is involved, the Alfv\'en wave driven model could reproduce the time-averaged radiative loss profile in the magnetic chromosphere. The time-averaged radiative loss profile is consistent with that in the classic atmospheric models. In addition, the energy transported to the corona could also meet the requirement of coronal heating in the quiet region, which is consistent with previous studies. However, the temperature in the magnetic chromosphere is apparently lower than that in the classic atmospheric model. Comparison with previous studies indicate that one needs to be careful when applying the simplified radiative loss term while studying chromospheric dynamics. \rv{For example, when quantifying spicule height and coronal energy flux, simplified radiative loss} will involve new artificial parameters \rv{which affect stratification in the chromosphere and further lead to change in spicule heights and the energy flux of waves.}

\rv{We acknowledge the referee for valuable comments. T. Y. is supported by JSPS
KAKENHI grant No. 15H03640.}

\appendix
\section{Selection of parameters in estimation of heating rate
  \label{app1}}
  \begin{figure}[h]%
\centering
\includegraphics[width=8cm]{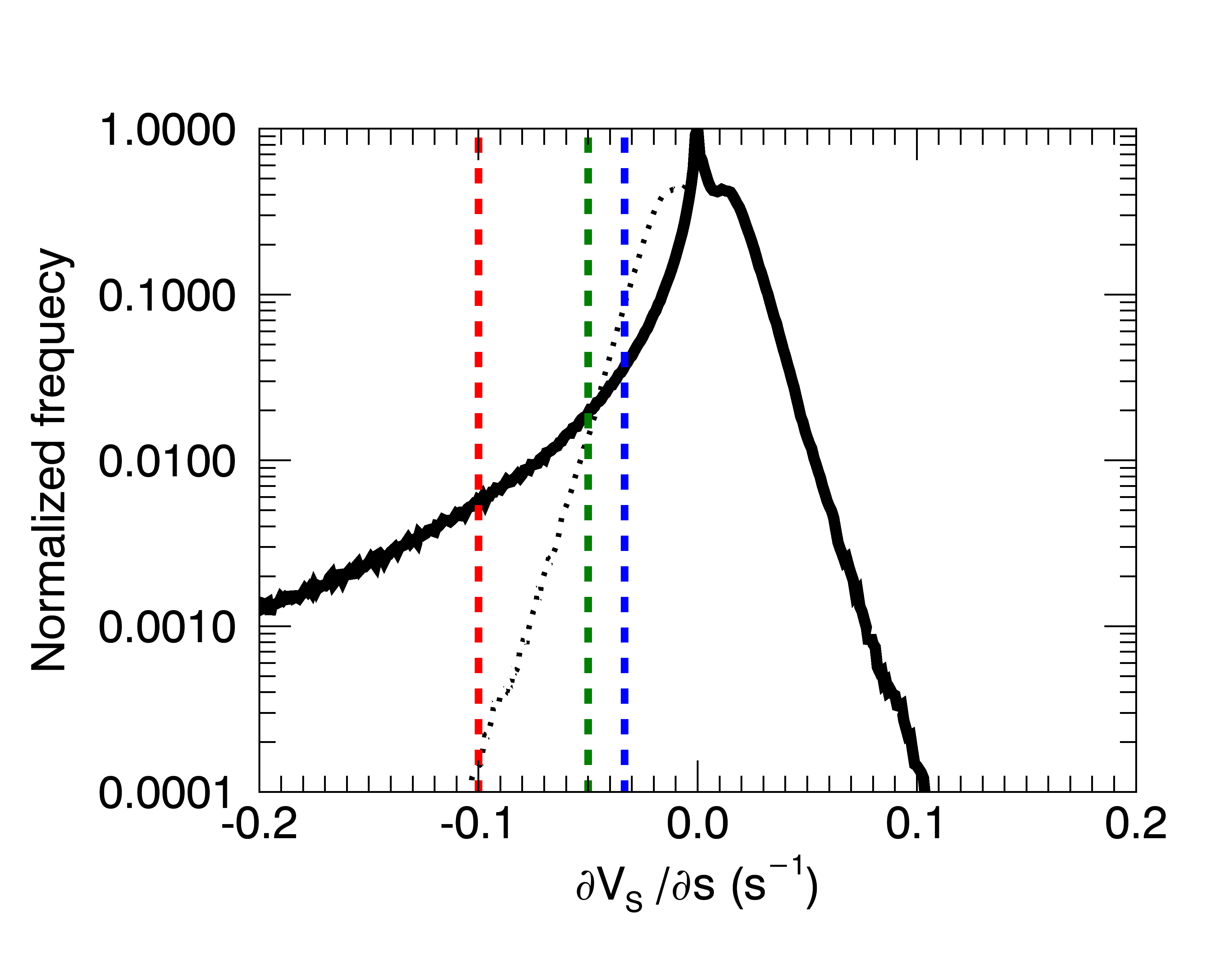}
\caption{Thick black line is normalized occurrence frequency distribution of $\partial V_s/\partial s$. The dotted black line is normalized occurrence frequency versus $-\partial V_s/\partial s$ for $\partial V_s/\partial s > 0$ (dotted black line and right part of the thick black line with $\partial V_s/\partial s > 0$ are symmetric with respect to $\partial V_s/\partial s = 0$). Vertical dashed red, green and blue lines are $\partial V_s/\partial s=$ 1/10 $\text{s}^{-1}$, 1/20 $\text{s}^{-1}$, and 1/30 $\text{s}^{-1}$ respectively.}

   \label{fig:freqdiv}
\end{figure}
For estimation of heating rate in our simulation, we need to identify each shock by divergence of velocity. We use $\frac{\partial {V_s}}{\partial s} \le -\frac{1}{t_{\text{c}}}$ to select shock regions (Equation \ref{eq:sel}) and heating rate is calculated using physical parameters in upstream and downstream regions of the shock wave (Equation \ref{eq:cal}) . $t_{\text{c}}$ in Equation (\ref{eq:sel}) is required to be large enough for including weaker shocks while small enough to exclude compression from linear propagation of waves. For this purpose, we plot the occurrence frequency distribution of $\partial V_s/\partial s$ in Figure \ref{fig:freqdiv}. For linear propagating waves, it is expected to have a symmetric distribution with respect to $\partial V_s/\partial s=0$. The actual distribution is asymmetric which has larger frequency for negative $\partial V_s/\partial s$ due to the formation of shocks. For distribution in $\partial V_s/\partial s>0$, we plot its symmetric part with respect to $\partial V_s/\partial s=0$ in dotted line. The actual frequency distribution is much larger than the symmetric part for the threshold $t_{\text{c}}=10$ s (red dotted line). We conclude that $t_{\text{c}}=10$ is small enough to exclude compression from the linear propagation of waves. We also apply $t_{\text{c}}=$ 20 s and 30 s for comparison. A larger threshold will include weaker shocks as well as the possibility of overestimation of heating rate because compression in linear propagating waves may be included. Figure \ref{fig:heating} shows that the heating rate above 1 Mm is similar for the three different thresholds, which indicates that the threshold $t_{\text{c}}=10$ s does give a good estimation of heating rate and we do not need to concern about an underestimation due to weak shocks that are excluded by this threshold.

In our calculation of heating rate, $w_{\text{shock}}$ is arbitrary that only affect the local spatial distribution of heating rate. For a single shock wave, as we set $Q_{\text{heat}}=0$ outside the shock region and the heating rate is constant in the shock region, the spatial integration of $Q_{\text{heat}}$ dose not depend on $w_{\text{shock}}$.

\bibliographystyle{apj}
\bibliography{reference}




\end{document}